\documentclass[twocolumn]{pasj00}

\begin{document}
\SetRunningHead{M. Shirahata et al.}{CO absorption toward IRAS~08572$+$3915}
\Received{2011/12/20}
\Accepted{2012/08/20}
\Published{2013/02/25}

\title{Infrared Spectroscopy of CO Ro-vibrational Absorption Lines toward the Obscured AGN IRAS~08572$+$3915
\thanks{Based on data collected at Subaru Telescope, which is 
operated by the National Astronomical Observatory of Japan.}}

\author{
  Mai \textsc{Shirahata}\altaffilmark{1,2}, 
  Takao \textsc{Nakagawa}\altaffilmark{1}, 
  Tomonori \textsc{Usuda}\altaffilmark{2}, 
  Miwa \textsc{Goto}\altaffilmark{3}, 
  Hiroshi \textsc{Suto}\altaffilmark{4}, 
  and
  T. R. Geballe\altaffilmark{5}}
\altaffiltext{1}{Institute of Space and Astronautical Science(ISAS), Japan Aerospace Exploration Agency(JAXA), 3-1-1 Yoshinodai, Chuo-ku, Sagamihara, Kanagawa 252-5210}
\email{sirahata@ir.isas.jaxa.jp}
\altaffiltext{2}{Subaru Telescope, 650 North A'ohoku Place, Hilo, HI 96720, U.S.A.}
\altaffiltext{3}{Universit\"ats-Sternwarte M\"unchen, Scheinerstr. 1, D-81679 Munich, Germany}
\altaffiltext{4}{National Astronomical Observatory of Japan, National Institutes of Natural Sciences, 2-21-1 Osawa, Mitaka, Tokyo 181-8588}
\altaffiltext{5}{Gemini Observatory, 670 North A'ohoku Place, Hilo, HI 96720, U.S.A.}


\maketitle

\begin{abstract}

We present high-resolution spectroscopy of gaseous CO absorption in 
the fundamental ro-vibrational band toward the heavily obscured active 
galactic nucleus (AGN) IRAS~08572$+$3915. We have detected 
absorption lines up to highly excited rotational levels ($J\leqq$ 17). 
The velocity profiles reveal three distinct components, the strongest 
and broadest ($\Delta v$ $>$ 200 km~s$^{-1}$) of 
which is due to blueshifted ($-$160 km~s$^{-1}$) gas at a temperature of $\sim$ 270 K 
absorbing at velocities as high as $-$400 km~s$^{-1}$. 
A much weaker but even warmer ($\sim$ 700 K) component, 
which is highly redshifted ($+$100 km~s$^{-1}$), 
is also detected, in addition to a cold ($\sim$ 20 K) component 
centered at the systemic velocity of the galaxy. 
On the assumption of local thermodynamic equilibrium, 
the column density of CO in the 270 K component is 
$N_{\rm{CO}}\sim4.5\times10^{18}$ $\rm{cm^{-2}}$, 
which in fully molecular gas corresponds to a $\rm{H_2}$ column density of 
$N_{\rm{H_2}}\sim2.5\times10^{22}$ $\rm{cm^{-2}}$. 
The thermal excitation of CO up to the observed high rotational levels 
requires a density greater than $n_c(\rm{H_2})>2\times10^{7}$ $\rm{cm^{-3}}$, 
implying that the thickness of the warm absorbing layer is 
extremely small ($\Delta d<4\times10^{-2}$ pc) even if it is highly clumped. 
The large column densities and high radial velocities 
associated with these warm components, as well as their temperatures, 
indicate that they originate in molecular clouds near the central engine of the AGN. 
\end{abstract}

\section{Introduction}\label{sec:intro}

Active galactic nuclei (AGNs) are broadly classified into two types: 
type 1 AGNs, which display broad permitted emission lines in the optical spectrum; 
and type 2 AGNs, which display both permitted and forbidden lines with narrow line widths. 
The detection of broad emission lines in polarized light from type 2 AGNs 
(Antonucci \& Miller 1985) present a clue to tie these two types together. 
The AGN activity is powered by a supermassive black hole ($10^6$--$10^9$ $\MO$) 
and its accretion disk, which extends to $\leq$1 pc. 
This central engine is surrounded by an optically and geometrically thick dusty torus, 
extending to $\sim$ 1--100 pc. 
In the ``unified model'' of AGNs (e.g., Antonucci 1993), 
it is postulated that much of the variety in types of AGNs is not due to 
intrinsic differences between them, but is largely the result of 
their strongly non-spherical geometries and the 
different orientations of their central regions with respect to our lines of sight. 
Specifically, in a type 1 AGN the torus is observed face-on and 
the interior may be observed directly, 
while in a type 2 AGN the torus is observed edge-on, which 
prevents a direct view of the nucleus. 
The obscuring torus around the central engine is the key element for the AGN unified model, 
and the strongest verification of it would come from the detection of the molecular torus itself. 
Recent observations at many wavelengths, for example the detection of 
silicate emission from type 1 AGNs (e.g., Hao et al., 2007), surely show 
the presence of obscuring tori surrounding the nuclei. 
However, the physical conditions and structure of the torus 
have never been measured directly, and we have only a crude idea of them. 
The direct measurements of the physical conditions (temperature and column density) 
of the molecular torus and the determination of the gas kinematics within the torus 
are long awaited goals of extragalactic astronomy. 

Carbon monoxide (CO) is the most abundant interstellar molecule after 
hydrogen molecule ($\rm{H_2}$). For the study of molecular clouds, 
pure rotational emission lines of CO have been extensively observed 
in the millimeter and sub-millimeter wavelength regions. 
However, to date such observations have not been suitable for the detailed characterization 
of distant molecular clouds around the AGNs, because they have provided information on 
neither the detailed spatial structure (due to the relatively 
large beam sizes of current millimeter telescopes) nor the physical 
conditions (due to the limited number of observable lines) of the clouds. 
In addition, millimeter observations can be contaminated by large-scale 
CO emission in the host galaxies on scales of 10--100 kpc, 
which can make it difficult to clearly distinguish the molecular gas 
in close proximity to the AGN, i.e. the putative molecular torus.

The technique we employ in the current work is high-resolution 
spectroscopy of absorption lines of the CO fundamental 
($v$=1$\leftarrow$0) ro-vibration band, whose center lies near 4.7 
$\rm{\mu m}$. Continuum emission associated with the bright, compact 
central region of an AGN is used as a background source and the foreground 
molecular clouds are to be observed in absorption. This technique 
achieves very high spatial resolution, since the resolution is 
determined by the size of the background continuum source. In addition, 
a large number of CO lines covering a wide range of rotational levels 
can be observed simultaneously with the same instrument under the same 
conditions. This enables accurate determination of the physical 
conditions, such as temperatures and column densities 
of the absorbing molecular clouds. 

In spite of these advantages, most previous ground-based observations 
of the 4.7 $\rm{\mu m}$ CO absorption spectrum have been limited to Galactic sources 
(e.g., Scoville et al. 1983; Mitchell \& Maillard 1993; Nakagawa et al. 
1997; Goto et al. 2003), with only a few examples 
toward extragalactic objects (Spoon et al. 2003; Geballe et al. 2006). 
This is mainly due to difficulties in achieving high sensitivities 
at the wavelength around 4.7 $\rm{\mu m}$. 
8-m class telescopes with high sensitivity, now are making it possible 
to observe the CO ro-vibrational absorption lines in 
molecular clouds around some AGNs. 

One class of AGNs whose central nuclei are obscured by gas and dust in 
our line of sight, the so-called Seyfert 2 galaxies, would appear to be 
the most suitable for such observations. However, spectra of objects in 
this class have to date shown no significant absorption by the CO 
fundamental ro-vibration band (Lutz et al. 2004), in some cases down to 
very low limits (Mason et al. 2006; Geballe et al. 2009). On the other 
hand, the {\it Spitzer Space Telescope} has detected a strong 
absorption by gaseous CO toward the dusty Ultra-Luminous InfraRed 
Galaxy (ULIRG), IRAS~F00183$-$7111 (Spoon et al. 2004). Observations at 
the VLT by Risaliti et al. (2006) and Sani et al. (2008) also have 
detected CO fundamental band absorption toward some ULIRGs whose central 
sources are heavily obscured. These results suggest that CO absorption 
does not appear in typical Seyfert 2 galaxies but can be found in at 
least some heavily obscured AGNs. The reasons for this are unclear, 
although in the case of the prototypical Seyfert 2 galaxy, NGC~1068, 
Geballe et al. (2009) have concluded that most of the gas in front of 
its AGN is diffuse in nature and therefore contains little CO. Both 
classes of objects, Seyfert 2 galaxies and ULIRGs, 
appear to have considerable obscuring material near their AGNs. 
However, hard X-ray observations (Franceschini et al. 2003; 
Ptak et al. 2003) suggested that AGNs in ULIRGs are not necessarily the 
dominant source of nuclear luminosity and that nuclear star formation 
is a significant contributor. Ongoing star formation in the nuclei of 
ULIRGs implies the presence of large amounts of molecular gas there. 

ULIRGs (Sanders \& Mirabel 1996) are galaxies whose extremely 
large, quasar-like luminosities ($L_{\rm{IR}} \geq 10^{12} \LO$) emerge 
at least 90 percent in the form of infrared dust emission. 
They are associated with interacting or merging gas-rich galaxies 
(Armus et al. 1987; Sanders et al. 1988; Leech et al. 1994; Clements et al. 1996; Murphy et al. 1996), 
in which the collision produces either an extreme burst of star formation 
(usually near the nucleus of one of the merging galaxies), 
a greatly increased rate of infall into the vicinity of an AGN, or both. 
In many cases it is difficult to distinguish between the two phenomena, 
because the large amount of dust prevents direct observation 
of the energy source. However, ULIRGs with bright pointlike nuclei 
at mid-infrared wavelengths, whose spectra show strong dust absorption, 
are generally thought to possess luminous buried AGNs 
(Imanishi, Dudley \& Maloney 2006). In the case of a ULIRG-producing merger 
involving a galaxy that contains an AGN, the survival of 
the molecular torus in the AGN is unclear. 
The torus could be disrupted by gas falling into the nucleus or by ejection of gas 
caused by the rapid increase in luminosity of the AGN precipitated by 
such infall, or simply by intense heating by the AGN alone. 

Infrared spectroscopy of ULIRGs is a possible way to investigate 
the energetic events in the central region of obscured AGNs. 
For such purpose, the observations of the CO fundamental band absorption toward ULIRGs 
with space telescopes, Spitzer and AKARI, are effective in providing some information 
(Spoon et al. 2005, Imanishi et al. 2008, Imanishi et al. 2010). 
However, the spectral resolutions of those facilities are relatively low 
($R$ $\sim$ 600 for the Spitzer/IRS, $R$ $\sim$ 100 for the AKARI/IRC) 
and are insufficient to resolve molecular ro-vibrational 
bands into individual transitions, let alone resolve velocity structure. 
Thus, in order to reveal the physical conditions of the circumnuclear region 
and the origin of the CO absorption, we have obtained high-resolution 
spectra toward heavily obscured AGNs, using ground-based telescopes. 

In this paper we report on the high resolution 4.90--5.13 $\rm{\mu m}$ 
spectrum toward the obscured AGN IRAS~08572$+$3915 NW, 
using the 8.2-m Subaru Telescope. 
The detection of CO absorption lines in the fundamental band toward this 
galaxy was reported by Geballe et al. (2006), using the United Kingdom 
3.8-m Infrared Telescope (UKIRT). Although the signal-to-noise ratio 
of their spectrum is modest, they detected broad and complex 
absorption line profiles. Their spectrum shows that the line profile 
are dominated by a very broad absorption feature, which extends from 
0 to $-$400 km~s$^{-1}$ and peaking near $-$160 km~s$^{-1}$ relative to 
the systemic velocity. They also detected a second velocity component 
close to the systemic velocity, and marginally detected a third component 
that is redshifted. Our new and improved spectrum covers 
a greater portion of the band and has a considerably higher signal-to-noise ratio. 
Using it we can clearly separate the CO absorption into the three components 
deduced from the earlier spectrum. The new spectrum can be used 
to better constrain the physical conditions in the absorbing molecular clouds. 

The characteristics of IRAS~08572$+$3915 are reviewed in the next section. 
The observations and the data reduction procedures are described in Section~\ref{sec:obs}. 
In Section~\ref{sec:results}, we present the observed spectrum and the line profiles. 
Temperatures and column densities of the absorbing gas are estimated in Section~\ref{sec:tmpcol} 
assuming local thermodynamic equilibrium (LTE). 
The origin of each velocity component of the absorbing gas is discussed in Section~\ref{sec:component}, 
and the line of sight dimensions of the warm molecular clouds is estimated in Section~\ref{sec:size}. 
The locations of the continuum emitting and absorbing regions are 
discussed in Section~\ref{sec:cont}. 
In Section~\ref{sec:discussion}, we discuss the relation between the warm molecular clouds 
we discovered and the putative AGN molecular torus. 
A summary is given in Section~\ref{sec:summary}.

\section{IRAS~08572$+$3915}\label{sec:08}

IRAS~08572$+$3915 is a warm ULIRG ($L_{\rm{IR(8-1000\mu m)}} = 1.21\times 10^{12}$ $L_{\odot}$, 
$f_{25}/f_{60}=0.23$), discovered by IRAS (Sanders et al. 1988). 
The redshift of this galaxy, determined from CO 
line emission at millimeter wavelengths is 0.0583 (Evans et al. 2002). 
The galaxy is optically classified as 
a Low-Ionization Nuclear Emission-line Region (LINER) galaxy by Veilleux et al. (1999). 
The {\it Hubble Space Telescope} (HST) NICMOS image shows double nuclei 
separated by 5.5{\mbox{$^{\prime\prime}$}} ($\sim$ 6.6 kpc), with the 
northwest nucleus overwhelmingly the more luminous (Scoville et al. 2000). 
Mid-infrared high-resolution images indicate that almost 
all the mid-infrared luminosity comes from a spatially 
highly compact source ($<0\farcs3$) in the northwest nucleus (Soifer et al. 2000). 
The mid-infrared 8--13 $\rm{\mu m}$ spectrum shows an extremely deep 
silicate absorption feature ($\tau_{\rm{9.7}}$ $\sim$ 5.2; 
Dudley \& Wynn-Williams 1997, $\tau_{\rm{9.7}}$ $\sim$ 4.2; Spoon et al. 2006, Spoon et al. 2007), 
and the near-infrared 3--4 $\rm{\mu m}$ spectrum is 
dominated by the 3.4 $\rm{\mu m}$ hydrocarbon absorption feature 
($\tau_{3.4}$ $\sim$ 0.9) with no evidence for emission by polycyclic aromatic 
hydrocarbons (PAHs) (Imanishi \& Dudley 2000; 
Mason et al. 2004; Imanishi, Dudley \& Maloney 2006; Geballe et al. 2009). 
On the basis of their spectra, 
Imanishi, Dudley \& Maloney (2006) argued that the energy source of IRAS~08572$+$3915 
is very compact and is surrounded by optically thick dust, as is expected 
for a buried AGN. On the other hand, IRAS~08572$+$3915 does not show any 
indication of AGN activity in X-rays (Risaliti et al. 2000) or at radio 
wavelengths (Smith, Lonsdale \& Lonsdale 1998). This implies that 
the AGN is deeply buried in gas and dust ($N_{\rm{H}} \gg 10^{24}$ $\rm{cm^{-2}}$), 
i.e., it is a Compton thick AGN.

\section{Observations and data reduction}\label{sec:obs}

\begin{figure}
  \begin{center}
    \FigureFile(80mm,80mm){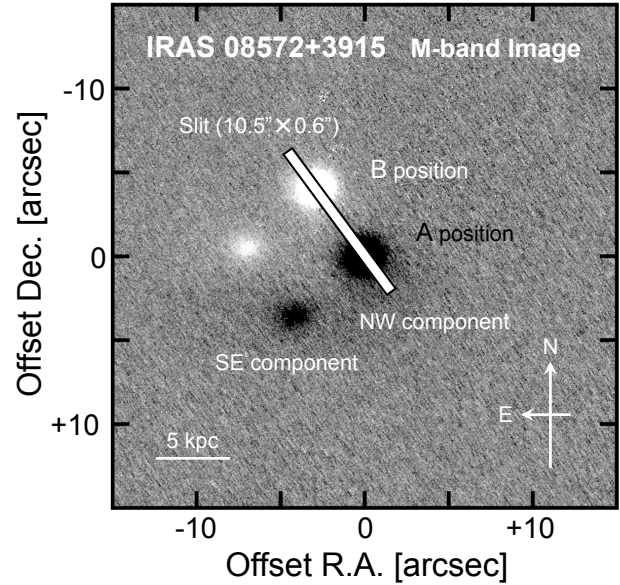}
  \end{center}
  \caption{
    Pair-subtracted $M$-band image of IRAS~08572$+$3915 with position of 
    slit indicated. The telescope was moved to alternately put the source 
    at the A and B positions in order to facilitate subtraction of the sky emission. 
    }\label{fig:SlitView}
\end{figure}

$M$-band spectra of the northwest nucleus (NW) of IRAS~08572$+$3915 were 
obtained using the Infrared Camera and Spectrograph (IRCS; Tokunaga et 
al. 1998; Kobayashi et al. 2000) on the 8.2 m Subaru Telescope 
atop Mauna Kea, Hawaii. The observations were performed on January 15 UT, 2004 
for the 5.00$-$5.13 $\mu$m spectrum and on February 3 UT, 2004 
for the 4.90$-$5.04 $\mu$m spectrum. Seeing conditions were 
0\farcs6$-$0\farcs8 FWHM in the $R$ and $I$ bands, in which we 
monitored the seeing, for both nights. An echelle and a cross-dispersing grating 
in the IRCS were employed to cover the following lines of 
the $^{12}$C$^{16}$O fundamental band in one exposure: 
$P$(7)$-$$P$(19) on January 15, $R$(3)$-$$R$(0), and $P$(1)$-$$P$(11) on February 3. 
A slit width of 0\farcs6 corresponding to a resolving power of 
$R$ $\sim$ 5,000 ($\Delta v$ $\sim$ 60 km~s$^{-1}$) was employed. 
The slit length was 10\farcs5. The slit position angle (P.A.) 
was set to 55\arcdeg\ east of north in order to observe only the 
northwest nucleus and avoid contamination from the southeast one. 
Figure~\ref{fig:SlitView} shows the slit position overlaid on the $M$-band 
image obtained with the slit viewer camera. The telescope was nodded 
5\farcs0 along the slit with the nodding frequency of $\sim$ 0.015 Hz 
for subtraction of the sky emission and the dark current. 
The on-source integration times were 64 and 44 minutes on 
January 15 and February 3, respectively. 

In order to cancel the telluric absorption, we obtained spectra of 
the early type standard stars HR~4534 (A3~V; $V$=2.14) and HR~3982 (B7~V; 
$V$=1.35) immediately after each observation with an airmass difference 
less than 0.15. We also obtained a spectroscopic flat field at the end of 
each night using the halogen lamp in the calibration unit in front of 
the entrance window of IRCS. 

\begin{figure*}
  \begin{center}
    \FigureFile(160mm,160mm){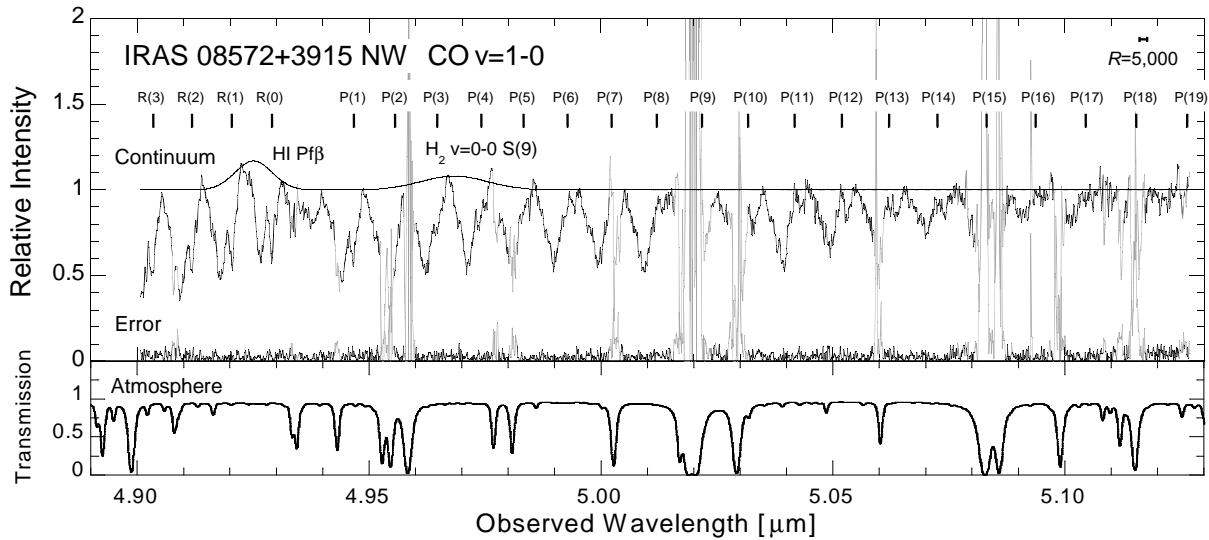}
  \end{center}
  \caption{
    Spectrum of IRAS~08572$+$3915 NW showing fundamental band ro-vibrational 
    lines of CO at 4.900--5.128 $\rm{\mu m}$. Data points shown in gray
    are excluded from the analysis, because of the low atmospheric 
    transmission ($<$ 0.5\%) or low signal to noise ratio 
    ($>$ 10\%). The adopted continuum is drawn as a thick line. The 
    wavelengths of the $R$- and $P$-branch lines of $\rm{^{12}C^{16}O}$ 
    corresponding to the redshift of the host galaxy are indicated on the 
    top by solid vertical lines. The error is estimated from the difference 
    between the spectra obtained at the A and B positions. The atmospheric 
    transmission curve was calculated with the ATRAN code (Lord 1992). 
    }\label{fig:Spectrum}
\end{figure*}

Standard data reduction procedures were employed. 
The observed spectral frames in the nodding sequence were 
coadded and differenced in order to remove the atmospheric emission. 
Flat-fielding was performed by dividing by the dark-subtracted lamp 
spectral image. Bad-pixels were identified on the basis of the 
statistics of the dark and flat field images, and removed by 
interpolation over surrounding pixels. One-dimensional spectra were 
extracted using the aperture extraction package in IRAF\footnote{IRAF 
(Image Reduction and Analysis Facility) is distributed by the National 
Optical Astronomy Observatories (NOAO) in Tucson, Arizona. NOAO is 
operated by the Association of Universities for Research in Astronomy 
(AURA), Inc., under cooperative agreement with the National Science 
Foundation.}. The emission from IRAS~08572$+$3915 NW was 
unresolved along the slit. The widths of the extraction apertures were adjusted to 
maximize the signal-to-noise ratios of the spectra and were 0\farcs 70 
on January 15 and 0\farcs 46 on February 3. Wavelength calibration was 
performed by comparing the positions of atmospheric absorption lines in 
observed spectra with the wavelengths of the lines in a model 
atmospheric transmission curve calculated using the ATRAN code (Lord 1992). 
The accuracy of the wavelength calibration is better than 6 km~s$^{-1}$.

\section{Results}\label{sec:results}

\begin{figure}
  \begin{center}
    \FigureFile(35mm,35mm){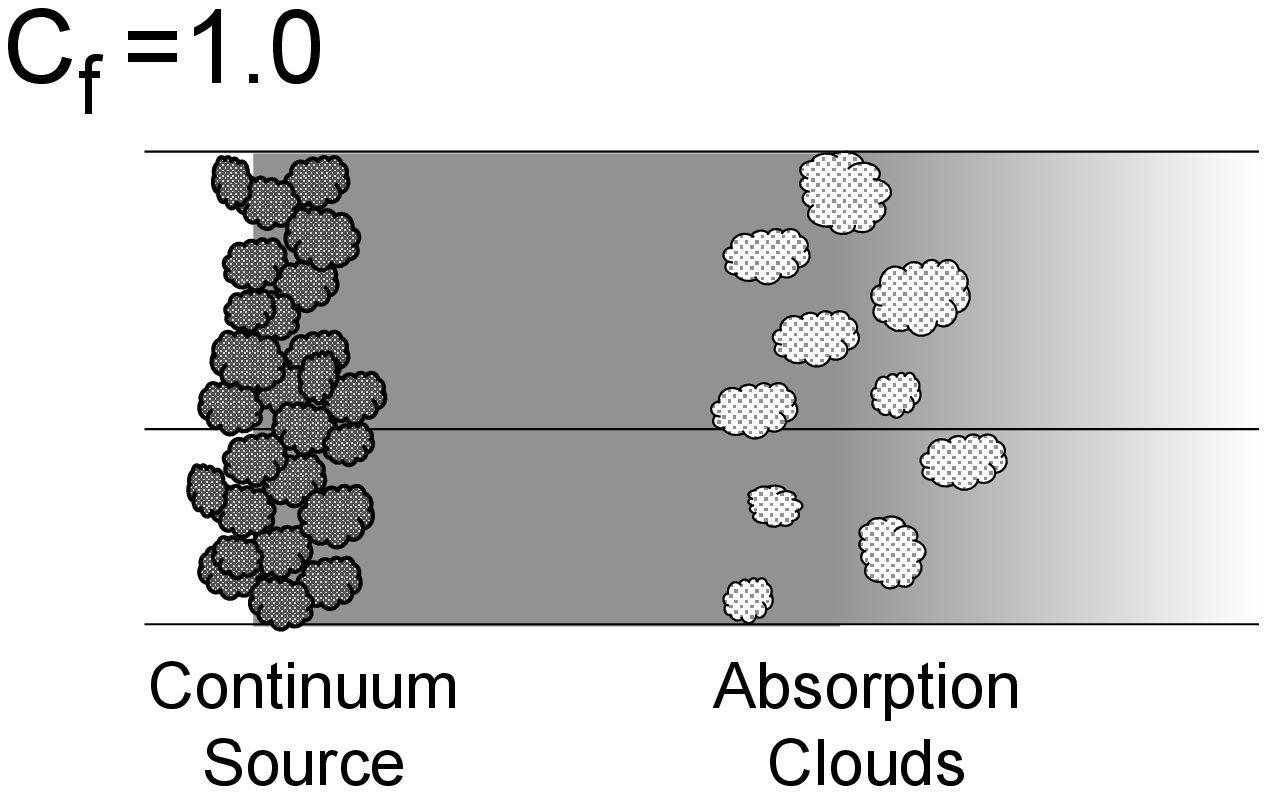} \qquad
    \FigureFile(35mm,35mm){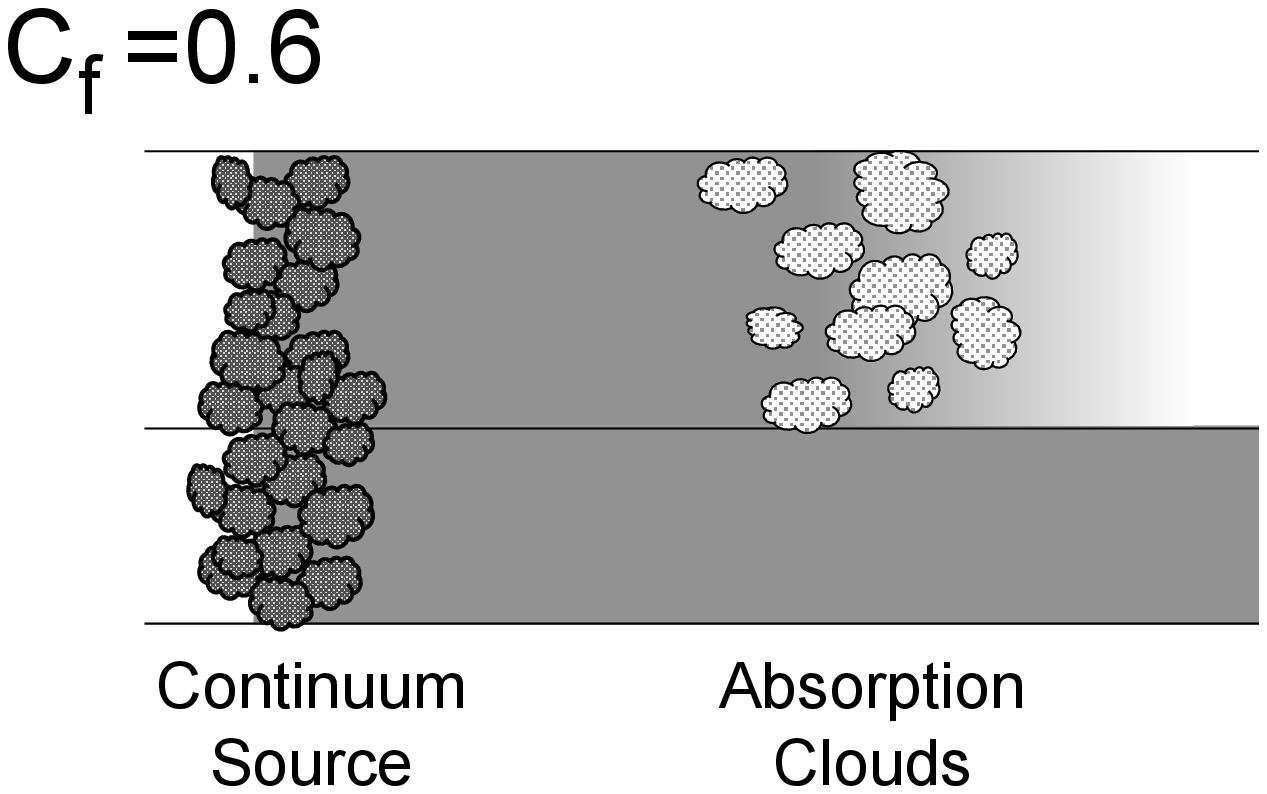}
  \end{center}
  \caption{
    Schematic view of a covering factor ($C_f$), the fraction of 
    the continuum source obscured by absorbing clouds along the line of sight. 
    (Left) covering factor is unity. (Right) covering factor is 0.6. 
    }\label{fig:schematic}
\end{figure}

The observed $M$-band spectrum of IRAS~08572$+$3915 NW is shown in 
Figure~\ref{fig:Spectrum} together with the ATRAN model spectrum of the 
atmospheric transmission. Absorption lines of CO in the $v$=0 state 
arising from rotational levels as high as $J$=17 were detected. 
These lines are extremely broad; 
the typical full widths at zero intensity (FWZIs) are more than 500 km~s$^{-1}$. 
The maximum depth of the absorption lines is about 60\% of the continuum level. 
This implies that the area covering factor ($C_f$; see Figure~\ref{fig:schematic}), 
the fraction of absorbed area in the line of sight, is at least 60\%.

Calculated wavelengths of the $R$- and $P$-branch lines of the 
$v$=1$\leftarrow$0 band of $\rm{^{12}C^{16}O}$, shifted to the 
redshift of the host galaxy, are denoted in Figure~\ref{fig:Spectrum} 
by solid vertical lines. The spacings of the observed lines agree well 
with the calculated ones. This wavelength region also contains 
absorption lines of isotopomers, $^{13}\rm{C}^{16}\rm{O}$, 
$^{12}\rm{C}^{18}\rm{O}$, etc., and of other vibrational transitions such as 
$v$=2$\leftarrow$1, $v$=3$\leftarrow$2, and so on. 
There is no indication that any of those bands are present. 
The strengths of the isotopic lines would be much less than those of the 
$v$=1$\leftarrow$0 lines of $^{12}\rm{C}^{16}\rm{O}$ for the standard 
isotope ratios (Sakamoto et al. 1997) unless the $^{12}\rm{C}^{16}\rm{O}$ lines 
are extremely heavily saturated. The spacings of lines of 
other isotopomers and other vibrational states of $^{12}\rm{C}^{16}\rm{O}$ 
do not match those of the observed lines. Therefore, we conclude that 
all the observed absorption lines are $v$=1$\leftarrow$0 lines of $^{12}\rm{C}^{16}\rm{O}$. 
In the following, we use CO for $^{12}\rm{C}^{16}\rm{O}$. 

\begin{figure}
  \begin{center}
    \FigureFile(70mm,70mm){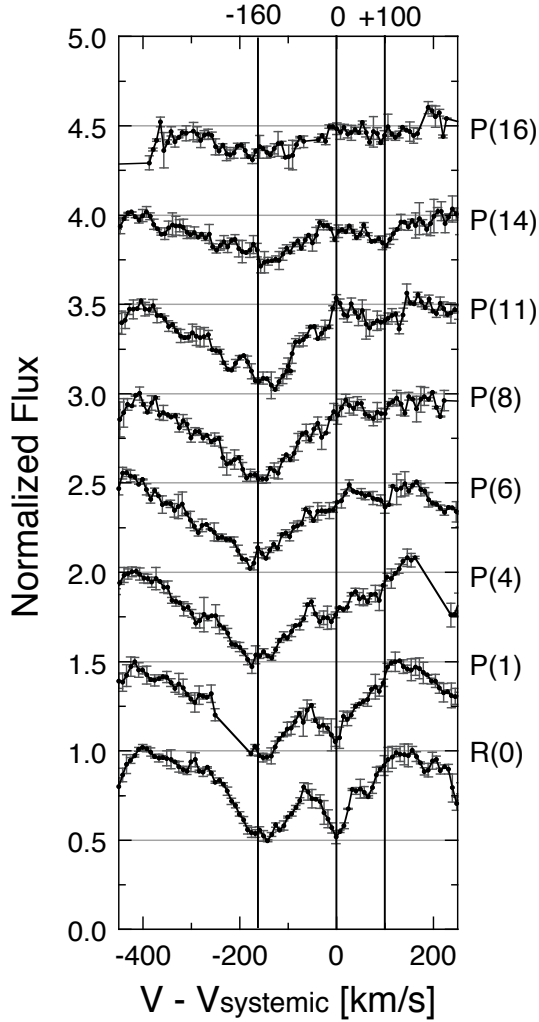}
  \end{center}
  \caption{
    Observed line profiles for various transitions in the spectrum shown in
    Figure~\ref{fig:Spectrum}. The horizontal axis is the velocity relative
    to the systemic velocity of the host galaxy. 
    }\label{fig:LineShape}
\end{figure}

Figure~\ref{fig:LineShape}, which shows the observed velocity profiles 
for various transitions, 
reveals remarkably broad and high-speed components. 
The velocity scale takes into account the orbital motion of the Earth 
relative to the Sun and the motion of the Sun relative to local standard 
of rest (Meeks et al. 1976). 
Although the line profiles are complex, they can be generally described as 
consisting of three velocity components: 
one centered at 0 km~s$^{-1}$ (the systemic velocity); 
and the others peaking at $-$160 km~s$^{-1}$ and $+$100 km~s$^{-1}$. 
The relative intensities, velocity extents, and 
degrees of overlap of these components with one another vary with 
rotational level. 
The component at the systemic velocity was detected in only the low-$J$ transitions 
($J\leqq5$), implying that the absorbing gas producing it has a relatively low temperature. 
The blueshifted component, the strongest of the three overall, 
was detected in all of the transitions including those at high-$J$, 
which indicates a significant contribution from high-temperature gas. 
Its line profiles are asymmetric and could consist of 
two or more velocity components. The redshifted component, which is weaker than the 
others, was detected only in high-$J$ transitions ($J\geqq4$), 
indicating that the temperature of the gas 
producing this component is higher than that producing the blueshifted component. 
IRAS~08572$+$3915 is the first extragalactic object in which 
the fundamental band of CO has been resolved into 
individual lines, and where the lines themselves are velocity-resolved.

\section{Physical conditions of molecular clouds}\label{sec:physical}

\subsection{Temperature and column density}\label{sec:tmpcol}

\subsubsection{Full coverage case}\label{sec:fullcoveragecase}

In this and the following subsection we analyze the spectrum in order to 
derive column densities and temperatures of the warm and cold components. 
In this subsection we make the simple assumption that 
the area covering factor ($C_f$) of the absorbing molecular gas is unity; 
in other words that for all transitions and at all velocities at 
which the CO absorbs, the absorbing gas completely covers the continuum source. 

In order to measure the intensities of the absorption lines, it is critical to 
know the continuum level accurately. Because the present 
spectrum is full of broad absorption lines, that level is difficult to be determined. 
We initially regard the local maxima of the continuum between each line as defining 
the continuum level. In that case the continuum flux density decreases 
gradually with increasing wavelength at wavelengths longer than 5.00 
$\rm{\mu m}$. We fit this portion of the continuum with a linear function. 
The apparent continuum on the short-wavelength side 
of the spectrum shows significant structure and is inconsistent with 
a linear dependence on wavelength. We attribute this 
to two emission lines: H\,{\small I} Pf$\beta$ and $\rm{H}_2$ 
$v$=0$\rightarrow$0 $S$(9) centered at 4.65378 $\rm{\mu m}$ and 
4.69462 $\rm{\mu m}$ in the rest frame, respectively. 
Goldader et al (1995) and Higdon et al (2006) have reported 
the detection of H\,{\small I} and $\rm{H}_2$ emission lines 
at near and mid-infrared wavelengths from this galaxy. 
The lower and upper level energies associated with the $\rm{H}_2$ $S$(9) 
line are approximately 4,600~K and 7,200~K, respectively and hence it 
is not surprising that the line is observed in emission. 
We assume that each of H\,{\small I} Pf$\beta$ and $\rm{H}_2$ $S$(9) lines 
have Gaussian profiles with the same velocity 
widths as H\,{\small I} Br$\gamma$ (241 km~s$^{-1}$) and $\rm{H}_2$ 
$v$=1$\rightarrow$0 $S$(1) (430 km~s$^{-1}$) observed by Goldader et al. (1995). 
The resulting best fit continuum spectrum synthesized from a straight 
line plus two Gaussian profiles is also plotted in Figure~\ref{fig:Spectrum}.
Our analysis of the few CO lines coincident 
in wavelength with these emission lines assumes that the emission lines 
are behind the absorbing CO. This appears to us to be the most likely 
situation for the Pf$\beta$ line, which arises in ionized gas. The 
$\rm{H}_2$ $S$(9) line occurs in molecular gas much hotter than the 
warm red- and blue-shifted gas producing CO absorption. It is possible, 
however, that it is physically associated with some of that gas and is 
in the foreground. However, the line is weak relative to the continuum 
and the continuum placement has only a small effect on the column 
densities of the one or two coincident CO lines and no effect on the 
principal results of this paper. 

The equivalent widths of the CO absorption lines, measured against the 
continuum level discussed above, were used to calculate the CO column 
density in the lower state of each transition. Since we found it 
impossible to definitively assign specific velocity intervals to the 
three velocity components discussed earlier, only total line equivalent 
widths $W_J$ across the entire profile were determined by 
numerical integration of the optical depth $\tau$, using 
\begin{equation}
  W_J = \int \frac{I_0-I_{\lambda}}{I_0}\ d \lambda = \frac{\lambda^2}{c} \int(1-e^{-\tau_{\nu}}) \ d \nu\ , \label{eq:EW}
\end{equation}
where $J$ is the rotational quantum number of the lower state, 
$I_0$ is the continuum level, $I_{\lambda}$ is the intensity of the spectrum, 
and $\lambda$ and $\nu$ is are wavelength and frequency of the line profile 
(but in the right portion of the equation $\lambda$ is the wavelength of the line center). 
When stimulated emission is negligible compared to absorption, 
the column density in the lower state of a transition $N_J$ is 
related to the equivalent width by 
\begin{equation}
  N_J = \frac{m_ec^2}{\pi e^2}\ \frac{1}{f \lambda^2}\ W_J \ , \label{eq:N} 
\end{equation}
in the optically thin case, 
where $\pi e^2/m_ec^2=8.8523\times10^{-13}$ cm, 
and $f$ is the oscillator strength of the transition. 
The oscillator strengths were taken from Goorvitch \& Chackerian (1994). 
The measured equivalent widths and column densities are summarized in Table~\ref{tab:boltzman08}. 

\begin{table*}
  \begin{center}
    {\scriptsize
      \caption{CO $v$=1$\leftarrow$0 Absorption Lines.}
      \label{tab:boltzman08}
      \begin{tabular}{lcccccccc}
        \hline
          &  &  &  & \multicolumn{2}{c}{Covering Factor 1} &  & \multicolumn{2}{c}{Covering Factor 0.6} \\ \cline{5-6} \cline{8-9}
        $\rm{^{12}C^{16}O}$ $v$=1$\leftarrow$0 & Wavelength & Energy & Oscillator strength & Equivalent width & Column density &  & Equivalent width & Column density \\
        Transition & [$\rm{\mu m}$] & $E_J/k$ [K] & f [$10^{-6}$] & $W_{\lambda}$ [$10^{-3}$ $\rm{\mu m}$] & $N_J$ [$10^{17}$ $\rm{cm^{-2}}$] &  & $W_{\lambda}$ [$10^{-3}$ $\rm{\mu m}$] & $N_J$ [$10^{17}$ $\rm{cm^{-2}}$] \\ \hline

        R(2)....................... & 4.6412 & 16.596 & 6.8760 & $3.42\pm0.04$ & $2.61\pm0.03$ & & ... & ...\\
        R(1)....................... & 4.6493 & 5.5321 & 7.6267 & $2.75\pm0.03$ & $1.89\pm0.02$ & & $9.11\pm0.05$ & $6.25\pm0.04$ \\
        R(0)....................... & 4.6575 & 0.0000 & 11.420 & $2.37\pm0.03$ & $1.08\pm0.01$ & & $5.54\pm0.03$ & $2.53\pm0.02$ \\
        P(1)....................... & 4.6742 & 5.5321 & 3.7900 & $2.84\pm0.04$ & $3.87\pm0.05$ & & $8.43\pm0.08$ & $11.50\pm0.11$ \\
        P(2)....................... & 4.6826 & 16.596 & 4.5400 & $2.86\pm0.10$ & $3.24\pm0.11$ & & $8.72\pm0.69$ & $9.89\pm0.78$ \\
        P(3)....................... & 4.6912 & 33.192 & 4.8543 & $2.98\pm0.10$ & $3.15\pm0.10$ & & $7.19\pm0.09$ & $7.60\pm0.10$ \\
        P(4)....................... & 4.7002 & 55.318 & 5.0233 & $2.40\pm0.03$ & $2.44\pm0.03$ & & $5.55\pm0.06$ & $5.65\pm0.06$ \\
        P(5)....................... & 4.7088 & 82.975 & 5.1273 & $2.00\pm0.04$ & $1.99\pm0.04$ & & $3.90\pm0.05$ & $3.87\pm0.05$ \\
        P(6)....................... & 4.7177 & 116.16 & 5.1962 & $1.96\pm0.03$ & $1.91\pm0.03$ & & $3.53\pm0.03$ & $3.45\pm0.03$ \\
        P(7)....................... & 4.7267 & 154.87 & 5.2433 & $2.09\pm0.12$ & $2.02\pm0.11$ & & $3.64\pm0.12$ & $3.51\pm0.11$ \\
        P(8)....................... & 4.7359 & 199.11 & 5.2765 & $2.31\pm0.04$ & $2.20\pm0.03$ & & $4.87\pm0.04$ & $4.64\pm0.03$ \\
        P(9)....................... & 4.7451 & 248.88 & 5.3000 & ... & ... & & ... & ... \\
        P(10)..................... & 4.7545 & 304.16 & 5.3190 & ... & ... & & ... & ... \\
        P(11)..................... & 4.7640 & 364.97 & 5.3304 & $1.90\pm0.03$ & $1.77\pm0.03$ & & $3.57\pm0.03$ & $3.34\pm0.03$ \\
        P(12)..................... & 4.7736 & 431.30 & 5.3400 & $1.56\pm0.04$ & $1.45\pm0.04$ & & $2.24\pm0.03$ & $2.08\pm0.03$\\
        P(13)..................... & 4.7833 & 503.14 & 5.3444 & $1.40\pm0.04$ & $1.29\pm0.04$ & & $1.75\pm0.04$ & $1.62\pm0.04$ \\
        P(14)..................... & 4.7931 & 580.49 & 5.3483 & $1.27\pm0.04$ & $1.17\pm0.03$ & & $1.56\pm0.04$ & $1.43\pm0.03$ \\
        P(15)..................... & 4.8031 & 663.35 & 5.3484 & ... & ... & & ... & ... \\
        P(16)..................... & 4.8131 & 751.72 & 5.3485 & $0.65\pm0.03$ & $0.59\pm0.03$ & & $0.74\pm0.03$ & $0.67\pm0.03$ \\
        P(17)..................... & 4.8233 & 845.60 & 5.3457 & $0.60\pm0.06$ & $0.55\pm0.05$ & & $0.72\pm0.06$ & $0.66\pm0.05$ \\
        P(18)..................... & 4.8336 & 944.97 & 5.3432 & $0.60\pm0.19$ & $0.54\pm0.17$ & & $0.78\pm0.18$ & $0.71\pm0.17$ \\ \hline \\

        \multicolumn{9}{@{}l@{}}{\hbox to 0pt{\parbox{180mm}{\footnotesize
              Notes. 
              \par\noindent
              Col.(3): Energy state of the lower level. 
              \par\noindent
              Col.(4): Oscillator strength given in Goorvitch \& Chackerian (1994). 
              \par\noindent
              Col.(5): Observed equivalent width, on the assumption covering factor of 1.0. 
              \par\noindent
              Col.(6): Column density of CO in the lower rotational state $J$, 
              on the assumption covering factor of 1.0. 
              \par\noindent
              Col.(7): Observed equivalent width, on the assumption covering factor of 0.6. 
              \par\noindent
              Col.(8): Column density of CO in the lower rotational state $J$,
              on the assumption covering factor of 0.6. 
            }\hss}}
        
      \end{tabular}
    }
  \end{center}
\end{table*}

On the assumption that the CO is in local thermodynamic equilibrium (LTE), 
the column density $N_J$ in the rotational level $J$ follows the Boltzmann distribution, 
\begin{eqnarray}
  N_J = \frac{N}{Z(T_{\rm{ex}})} \ (2J+1) \ \exp \left(-\frac{E_J}{kT_{\rm{ex}}} \right) \ ,\label{eq:boltzmann} 
\end{eqnarray}
where $N$ is the total column density, $T_{\rm{ex}}$ is 
the excitation temperature, $Z(T_{\rm{ex}})$ is the partition function 
which can be calculated by a polynomial interpolation provided in the 
HITRAN'08 database (Rothman et al. 2009), and $E_J\approx B_0 J(J+1)$ is 
the energy of the $J$ level where $B_0$ is the rotational constant: $B_0/k=2.765$ K. 
It is useful to plot the population diagram, 
$\ln [N_J/(2J+1)]$ versus $E_J/k$. If LTE is valid, the data points should 
be fit by one or more straight lines whose slopes are $1/T_{\rm{ex}}$. 

\begin{figure}
  \begin{center}
    \FigureFile(80mm,80mm){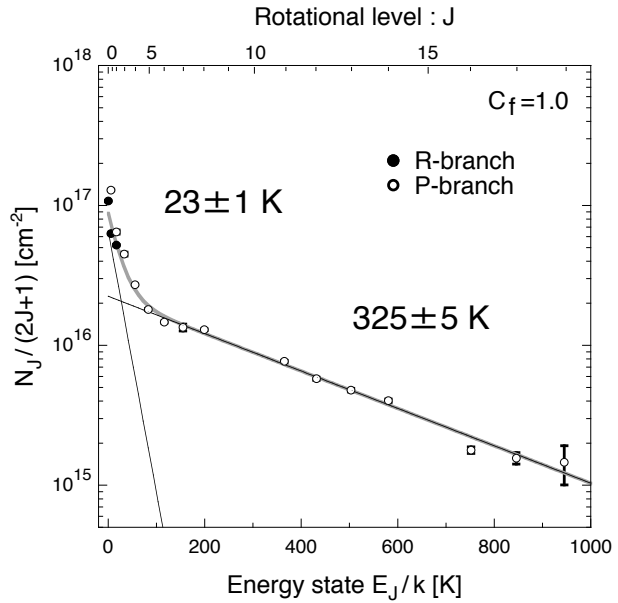}
  \end{center}
  \caption{
    Population diagram of CO absorption lines toward IRAS~08572$+$3915~NW 
    (see text) for unity covering factor. Temperatures shown correspond to the
    slopes of the two straight lines. Error bars are statistical
    uncertainties (1$\sigma$), and do not include systematic errors. 
    }\label{fig:BoltzmanC1}
\end{figure}

Figure~\ref{fig:BoltzmanC1} shows the population diagram for the 
observed CO lines (where $N$ is determined over the full velocity extent 
of each line). The diagram indicates that there are at least two components 
with different excitation temperatures. The excitation temperatures, 
determined by least-squares fits, are $T_{\rm{ex,cold}}=23\pm1$ K and 
$T_{\rm{ex,warm}}=325\pm5$ K. The CO column densities of these components 
calculated from Equation~\ref{eq:boltzmann} are 
$N_{\rm{CO,cold}}=(5.74\pm0.06) \times 10^{17}$ $\rm{cm^{-2}}$ and 
$N_{\rm{CO,warm}}=(2.65\pm0.04) \times 10^{18}$ $\rm{cm^{-2}}$ for 23 K 
and 325 K components, respectively. Assuming the standard cosmic abundance ratio 
$N_{\rm{CO}}/N_{\rm{H_2}} \sim 1.8\times10^{-4}$ in dense molecular clouds (Dickman 1978), 
these values correspond to 
$N_{\rm{H_2,cold}}=(3.19\pm0.03) \times 10^{21}$ $\rm{cm^{-2}}$ and 
$N_{\rm{H_2,warm}}=(1.47\pm0.02) \times 10^{22}$ $\rm{cm^{-2}}$, respectively. 
The derived temperature of the warm component is far higher than that found 
toward Galactic interstellar medium lines of sight (Lutz et al. 1996), 
and similar to that observed toward the obscured ULIRG IRAS~F00183$-$7111 by Spoon et al. (2004). 

\subsubsection{Partial coverage case}\label{sec:partcoveragecase}

As discussed later in Section~\ref{sec:cont}, the 5~$\rm{\mu m}$ 
continuum probably originates in dust at characteristic 
nuclear distances of a pc-scale. Krolik \& Begelman (1988) 
concluded that when a dusty molecular torus of such dimensions is 
associated with an AGN the torus is composed of a large number of small 
high-density molecular clumps. The geometry of the obscuring material 
in front of the nuclear continuum source in IRAS~08572$+$3915 is unclear, 
but may also consist of molecular clumps. Hence partial coverage of the 
continuum source should also be considered. 

Although in the previous subsection we derived the physical parameters 
of the gas clouds assuming that the covering factor is unity, 
that assumption may not be appropriate for the following reason. 
Generally speaking, analysis of the $R$- and $P$-branch transitions 
which originate from the same $J$ level should give the same column 
density of $N_J$. However, on the assumption of a covering factor of unity, 
the column densities calculated for the $R$-branch at $J$=1 and $J$=2 
are systematically lower than those for the $P$-branch. On the other hand, 
the oscillator strengths of the $R$-branch are systematically larger 
than those of the $P$-branch. The most probable explanation for this mismatch 
is that the absorption lines of the $R$-branch ($R$(1), $R$(2)) are 
saturated but the absorbing gas does not fully cover the 
continuum emitting region, so that some unobscured continuum is 
observed; i.e., the covering factor $C_f$ in the line of sight is less than unity. 

\begin{figure}
  \begin{center}
    \FigureFile(80mm,80mm){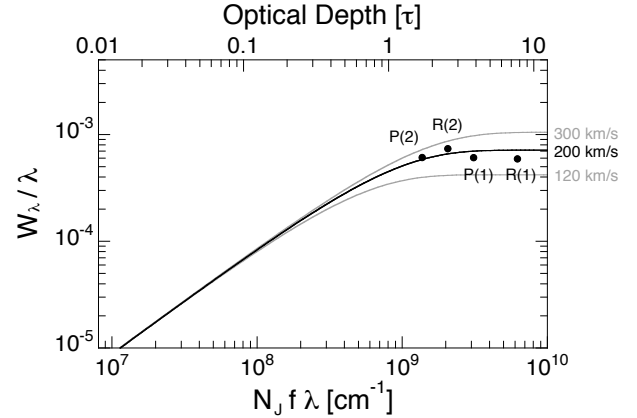}
  \end{center}
  \caption{
    Curves of growth assuming that an intrinsic line profile 
    is a Gaussian function. Three lines are corresponding to the line full 
    width half maxima (FWHMs) of 300, 200, and 120 km~s$^{-1}$. 
    A good match is obtained for FWHM $=$ 200 km~s$^{-1}$, and the 
    peak optical depth in the case of FWHM $=$ 200 km~s$^{-1}$ 
    can be read off of the upper $x$-axis. 
    }\label{fig:Curveofgrowth}
\end{figure}

The optical depths of the low-$J$ lines can be estimated by 
the curve-of-growth method. In Figure~\ref{fig:Curveofgrowth}, 
we show curves of growth, assuming for simplicity that 
an intrinsic line profile can be represented by a Gaussian function. 
Three curves of growth are shown, for different velocity widths 
(FWHM $=$ $2\sigma\sqrt{2\ln2}$ $=$ 300, 200, 120 km~s$^{-1}$; 
where $\sigma$ is the dispersion of a Gaussian function). 
Over-plotted on the figure are 
the absorption-line data from Table~\ref{tab:boltzman08}. 
In the optically thin part of the curve 
the equivalent width is proportional to the column density, 
while the equivalent width saturates in the optically thick part. 
For each absorption line plotted, 
the vertical position is determined from the observed equivalent width. 
For each absorption line pair from the same $J$ level, 
i.e., $R$(1) and $P$(1), $R$(2) and $P$(2), 
the horizontal positions have been shifted along the $x$-axis, 
while keeping their ratio of $f\lambda$ values fixed, 
until the pair match a growth curve. 
Their final positions give the column density for rotational level $J$. 
The measured equivalent widths best match the 
model curve with FWHM $=$ 200 km~s$^{-1}$. 
The corresponding optical depth of the gaussian profile 
is calculated analytically. 

Using this method the optical depths obtained for the $J$=1 and $J$=2 levels of CO 
are $\tau_{R(1)}\sim7.8$ and $\tau_{P(1)}\sim3.9$, and $\tau_{R(2)}\sim2.6$ 
and $\tau_{P(2)}\sim1.7$, respectively. These results suggest that the low-$J$ lines are 
heavily saturated and thus that the covering factor ($C_f$) of the 
absorbing molecular clouds is closer to 0.6 than 1.0. Ideally, one should 
use the curve-of-growth analysis method for all absorption lines. 
However, this was not possible for $J\geqq3$ lines 
since their $R$-branch lines were outside our wavelength coverage. 
We therefore applied $C_f=0.6$ to all the lines, and derived 
the column density of each line by the corrected optical depth of 
\begin{equation}
  \tau_{\nu} = - \ln \left( \frac{I_{\nu}+C_f-1}{C_f} \right). 
\end{equation} 
The equivalent widths and column densities of each transition line 
on the assumption of $C_f=0.6$ are summarized in Table~\ref{tab:boltzman08}. 
For low-$J$ lines, the column densities for $C_f=0.6$ are $2-4$ times 
larger than those for $C_f=1.0$, while 
for high-$J$ lines, the column densities for $C_f=0.6$ are slightly ($<2$ times) 
larger than those for $C_f=1.0$. 

\begin{figure}
  \begin{center}
    \FigureFile(80mm,80mm){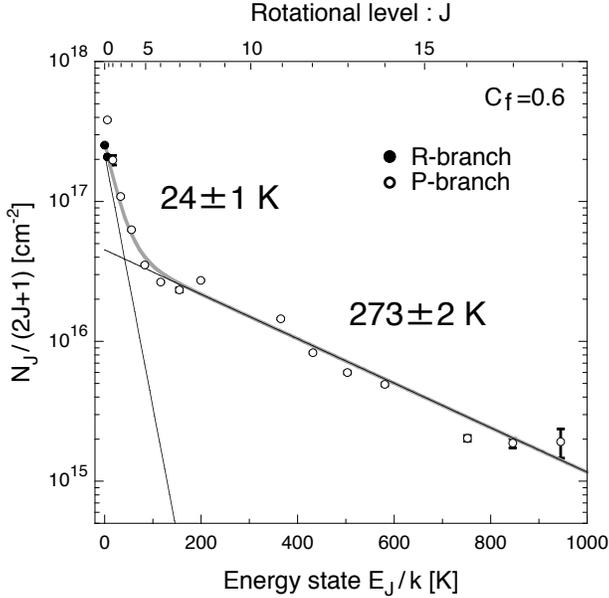}
  \end{center}
  \caption{
    Same as Figure~\ref{fig:BoltzmanC1}, but assuming a covering factor of 0.6.  
    }\label{fig:BoltzmanC6}
\end{figure}

Figure~\ref{fig:BoltzmanC6} shows the population diagram obtained on 
the assumption that $C_f=0.6$; it should be compared with Figure~\ref{fig:BoltzmanC1}. 
The derived temperatures and column densities for both cases 
are summarized in Table~\ref{tab:08_summary}. For the cold component, 
the excitation temperature is $T_{\rm{ex,cold}}=24\pm1$ K, 
which is similar to the $C_f=1.0$ case. The derived column density is 
$N_{\rm{CO,cold}}=(1.98\pm0.04) \times 10^{18}$ $\rm{cm^{-2}}$, 
which is three times larger than that of the $C_f=1.0$ case. 
On the other hand, for the warm component the excitation temperature is 
$T_{\rm{ex,warm}}=273\pm2$ K, which is slightly lower than that of the 
$C_f=1.0$ case, and the column density is 
$N_{\rm{CO,warm}}=(4.48\pm0.04) \times 10^{18}$ $\rm{cm^{-2}}$, 
which is slightly larger than that of the $C_f=1.0$ case. 
As a result, the corresponding $\rm{H_2}$ column densities are 
$N_{\rm{H_2,cold}}=(1.10\pm0.01) \times 10^{22}$ $\rm{cm^{-2}}$ and 
$N_{\rm{H_2,warm}}=(2.49\pm0.02) \times 10^{22}$ $\rm{cm^{-2}}$ 
for 24 K and 273 K components, respectively, 
when the standard value of [CO]/[H$_2$] in dense clouds is adopted. 

\begin{table*}
  \begin{center}
    \caption{Summary of temperatures and column densities.}
    \label{tab:08_summary}
    \begin{tabular}{lcccc}
      \hline
      Covering factor & Component & Temperature & CO column density & $\rm{H_2}$ column density \\
        &  & (K) & ($10^{18}$ $\rm{cm^{-2}}$) & ($10^{22}$ $\rm{cm^{-2}}$) \\ \hline

        $C_f=1.0$...............& Cold & $23\pm1$ & $0.574\pm0.006$ & $0.319\pm0.003$ \\
        $C_f=1.0$...............& Warm & $325\pm5$ & $2.65\pm0.04$ & $1.47\pm0.02$ \\
        $C_f=0.6$...............& Cold & $24\pm1$ & $1.98\pm0.01$ & $1.10\pm0.01$ \\
        $C_f=0.6$...............& Warm & $273\pm2$ & $4.48\pm0.04$ & $2.49\pm0.02$ \\ \hline \\

        \multicolumn{5}{@{}l@{}}{\hbox to 0pt{\parbox{180mm}{\footnotesize
              Notes. 
              \par\noindent
              Col.(4): Total CO column density derived on the basis of the temperature in Col.(3). 
              \par\noindent
              Col.(5): Total $\rm{H_2}$ column density, 
              assuming the standard cosmic abundance ratio of $N_{\rm{CO}}/N_{\rm{H_2}}\sim1.8\times10^{-4}$ (Dickman 1978). 
            }\hss}}

      \end{tabular}
  \end{center}
\end{table*}

We caution that because the CO absorbs over an extremely wide velocity 
range the assumption that $C_f$ is independent of velocity, based only 
on analysis of the low-$J$ lines, whose equivalent widths have 
comparatively large contributions from the cold gas, is questionable. 
It also is not obvious that $C_f=0.6$ can be applied to all the lines. 
Thus there is some uncertainty in the estimates of the column density. 
Note, however, that the difference between the results with $C_f=0.6$ 
and 1.0 is not large for the warm component, which is the main topic of 
the present paper. Hence, we take $C_f=0.6$ as the nominal value for 
the remainder of the paper.

\subsection{Origin of each velocity component}\label{sec:component}

Our observations have clearly shown the existence in front of the 
northwestern nucleus of IRAS~08572$+$3915 of 
three basic gaseous components associated with different velocity ranges 
centered at the systemic velocity and centered at $-$160 km~s$^{-1}$ and 
$+$100 km~s$^{-1}$ relative to it, and 
with three widely different excitation temperatures, 
only two of which are apparent in Figure~\ref{fig:BoltzmanC6}. 
In this subsection we discuss the origin of each velocity component. 
We have not been successful at cleanly isolating these components 
in velocity space for detailed analysis and comparison, 
using standard deconvolution techniques, 
because the velocity profiles are too complicated. 
The velocity extents, velocity ranges, and degrees of overlap of the components 
vary with rotational level. Moreover, the profile of each velocity component 
is asymmetric, and each could consist of several sub-components. 
Nevertheless, the large temperature difference of the two dominant absorption 
components allows us to use Figure~\ref{fig:LineShape} and Figure~\ref{fig:BoltzmanC6} 
to crudely separate and simply characterize them. The much weaker third component 
differs sufficiently in temperature and velocity from the others 
that it can also be crudely characterized. 

The cold component at $\sim$24 K revealed by Figure~\ref{fig:BoltzmanC6} 
must correspond to CO absorption at velocities near the systemic velocity. 
As seen in Figure~\ref{fig:LineShape}, 
this component was detected only in transitions from levels with $J\leqq5$, 
implying a gas temperature around 30 K. 
We interpret this component as arising in interstellar clouds 
in the host galaxy located far from the nucleus, since its excitation temperature 
is low and is similar to those of giant molecular clouds in the Galaxy. 
The presence of a strong 3.4 $\mu$m carbonaceous dust absorption feature 
toward IRAS~08572$+$3915 (e.g., Imanishi, Dudley \& Maloney 2006) suggests 
that a considerable fraction of the interstellar molecular gas 
along the line of sight is in diffuse clouds, as is the case 
toward the Galactic center (Whittet et al. 1997). 
In diffuse clouds almost all of the carbon ($\sim$ 99\%) is atomic and 
there very little of it is in CO ($\sim$ 1\%). 
Thus, if the gas traced by the observed cold CO component is 
entirely in diffuse clouds, the $\rm{H_2}$ column density could 
be two orders of magnitude larger than the previously estimated value. 
However, the present CO data do not in themselves allow us to 
determine the origin of cold CO at the systemic velocity. 
Accordingly, the $\rm{H_2}$ column density of 
$1 \times 10^{22}$ $\rm{cm^{-2}}$ derived above is a lower limit. 

Similarly, the warm component at $\sim$ 273 K, 
which is the dominant contributor to Figure~\ref{fig:BoltzmanC6}, 
must be directly associated with the strongest absorption component, 
centered at $-$160 km~s$^{-1}$ in Figure~\ref{fig:LineShape}. 
This component was detected in all of the transitions and is strongest at $J$=3--5, 
implying a gas temperature around 200 K. 
Outside of the Galactic center, the temperatures of molecular clouds 
do not exceed 100 K (Dickman \& Clemens 1983), except in very limited volumes 
such as those close to the luminous stars and shocked regions. 
It thus seems unlikely this large column density of warm negative velocity gas 
is located outside of the nucleus of IRAS~08572$+$3915 NW. 
We propose that this component is close to the nucleus 
and is heated by the central engine of the AGN. 
Meijerink \& Spaans (2005) showed that molecular clouds exposed to 
X-ray radiation (X-ray dissociation regions) 
can be heated to temperatures such as those observed here 
(see also Maloney et al. 1996; Meijerink, Spaans \& Israel 2006). 
The warm molecular gas may also be collisionally 
heated due to interactions of portions of the gas having different velocities. 
However, although such turbulent heating is a possibility, 
the negative velocities of this warm component indicate that 
the bulk gas motion is outward from the nucleus. 

The weak $+$100 km~s$^{-1}$ component seen in Figure~\ref{fig:LineShape} 
does not contribute significantly to Figure~\ref{fig:BoltzmanC6} 
because of its small column density. 
The equivalent widths of this component in the $P(6)$ and $P(16)$ 
line profiles are almost the same. From them we estimate that 
the temperature of the redshifted component is around 700 K 
and that the column density of CO is $\sim10^{17}$ $\rm{cm^{-2}}$. 
If the source of the heating of this cloud is the central engine, 
then the redshifted component would be expected to arise from clouds 
which lie closer to the central engine than those producing the blueshifted component. 
Outflows have been observed toward various types of AGN, 
but there are few examples of infall (e.g., Crenshaw et al. 2003; M\"uller S\'anchez et al. 2009). 
We suggest that this velocity component traces the material fueling the central engine.

\subsection{Densities and thicknesses of the warm clouds}\label{sec:size}

The geometrical thickness of the absorbing cloud along the line of sight 
can be estimated from the column density derived above, if the gas 
particle density is known. As seen in the population diagram of 
Figure~\ref{fig:BoltzmanC6}, the warm ($\sim$ 270 K) component is well 
thermalized at least to the $J$=17 level. This indicates that 
the population distribution of the CO molecules in the rotational states 
is controlled by collisions with other species, presumably mostly $\rm{H_2}$. 

A high gas density is required to thermalize CO up to the $J$=17 level. 
The density at which collisional de-excitation competes with radiative 
de-excitation (the critical density; $n_c(\rm{H_2})$) for CO $J\to J-1$ 
is approximately $n_c(\rm{H_2})=4\times10^3$ $J^3$ $\rm{cm^{-3}}$, as 
described in Kramer et al. (2004). The resulting value is 
$n_{c(17\to16)}(\rm{H_2})=2\times10^7$ $\rm{cm^{-3}}$. Collisional 
transitions with $\Delta J >1$ are much less likely than $\Delta J=1$ 
transitions and hence the critical density is approximately the above value. 
This lower limit to the actual density is extremely high compared 
to the densities of typical molecular clouds in the Galaxy ($n\sim10^3$ $\rm{cm^{-3}}$). 
Such densities are, however, found in 
post-shock regions of molecular clouds, where molecular outflows from 
young stars collide with ambient molecular gas (e.g., Burton et al. 
1988; Vannier et al. 2001), and may also occur where molecular clouds collide. 

Assuming that the absorbing cloud is homogenous, 
the thickness of the absorbing layer, $\Delta d$, is 
\begin{eqnarray}
  \Delta d < \frac{ N_{\rm{H_2}} }{ n_c(\rm{H_2}) }\ \approx \ 4\times10^{-4} \ \rm{pc}. 
\end{eqnarray} 
The absorbing gas column thus appears to be extremely short along the line of sight. 
In general, dense molecular clouds are expected be highly clumpy, 
as a natural result of the interplay between the energy generated by 
self-gravity of the gas and radiative cooling. Clumpiness increases the 
size of molecular clouds with respect to the homogeneous case estimated above. 
It is rather difficult to estimate the volume filling factor, 
the fraction of absorbed volume in the beam area. Assuming that it 
is about 1--10\%, similar to the values in highly 
clumped clouds in the Galaxy (Genzel et al. 1985), the thickness of 
the absorbing molecular clouds would be 10--100 times greater; 
i.e., 0.004--0.04 pc. This is still much 
less than the thicknesses of molecular tori assumed in previous studies 
of AGN (e.g., Jaffe et al. 2004; Wada \& Tomisaka 2005).

\subsection{Extinction estimates and locations of emitters and absorbers}
\label{sec:cont}

As described in Section~\ref{sec:tmpcol}, we have detected 
the Pf$\beta$ hydrogen recombination line. 
The intensity ratios of hydrogen recombination lines 
can be used to estimate the extinction by foreground dust, 
assuming ``Menzel case {\it B}'' conditions (Osterbrock \& Ferland 2005), 
for typical of H\,{\small II} regions. 
We thus compare the observed flux of Pf$\beta$ 
(4.6538 $\rm{\mu m}$, $(3.2\pm0.5)\times10^{-17}$ $\rm{W/m^2}$), with 
that of Br$\gamma$ (2.1661 $\rm{\mu m}$, $(0.7\pm0.2)\times10^{-18}$ $\rm{W/m^2}$) 
observed by Goldader et al. (1995). The flux ratio Pf$\beta$/Br$\gamma$ is 
$\sim46$, whereas case {\it B} recombination theory predicts a ratio of 0.58. 
The difference suggests a differential extinction of $E(M-K)=4.75$ mag 
to the recombination line emitting region. 
According to the standard extinction curve model for Milky Way dust (Draine 2003), 
this value corresponds to $A_V=57$ mag, implying a H column density of 
$N_{\rm{H}}=1.07\times10^{23}$ $\rm{cm^{-2}}$ 
(using $N\rm{(H)}/A_{V}=$ $1.9\times10^{21}$ $\rm{cm^2~mag^{-1}}$; Bohlin et al. 1978). 
The ratio could be considerably underestimated because the aperture used for 
our Pf$\beta$ observation (0\farcs60 $\times$ 0\farcs46) is 100 times 
smaller than that for Br$\gamma$ (3\farcs0 $\times$ 9\farcs0). 
Thus, the above column density is a lower limit. 
Case {\it B} conditions are found in H\,{\small II} regions, 
but not in dense ionized winds; we caution that the physical conditions in the gas 
producing the hydrogen recombination lines is unknown. 

From Section~\ref{sec:partcoveragecase} the $\rm{H_2}$ column density 
summed from all CO velocity components is $N_{\rm{H_2}}\sim4\times10^{22}$ $\rm{cm^{-2}}$, 
slightly lower than the lower limit estimated from the H recombination lines. 
However, this estimate of $N(\rm{H_2}$) is itself a lower limit in that, 
as discussed earlier, some or all of the cold CO may reside in diffuse gas. 
X-ray observations toward IRAS~08572$+$3915 imply that 
the H column density to the AGN exceeds $N_{\rm{H}}\gg10^{24}$ $\rm{cm^{-2}}$ (Risaliti et al. 2000). 
For comparison, the total hydrogen column density estimated from 
9.7 $\rm{\mu m}$ silicate dust absorption is $N_{\rm{H}}\sim1\times10^{23}$ $\rm{cm^{-2}}$ 
($\tau_{\rm{9.7}}/A_V=$ 0.05--0.1; Roche \& Aitken 1984, 1985; 
$\tau_{\rm{9.7}}=4.2$; Spoon et al. 2006), comparable to 
the lower limit to the $\rm{H_2}$ column density derived from the CO absorption. 
We conclude that the column density of hydrogen in front of the AGN, 
although highly uncertain, exceeds $10^{23}$ $\rm{cm^{-2}}$.

Due to the large uncertainties in the above estimates, 
we cannot confidently assign relative locations along 
the line of site to the X-ray emitting region, recombination line emitting, 
infrared continuum emitting, and warm molecular absorbing regions 
based solely on estimates of foreground column densities of hydrogen. 
On the basis of energetics, however, 
one expects that the X-ray emitting gas must be interior to 
the recombination line emitting region. Both of those must be interior to 
the infrared continuum emitting regions at 5 and 10 microns, 
which themselves must be internal to the CO and silicate absorptions, respectively. 
These assignments are in fact consistent with the column densities derived above 
if they are actual values rather than (in some cases) lower limits. 
Also these suggest that the infrared continuum emission comes, 
not from the central engine itself, but from a location somewhat further from the engine. 
One expects that the 
continuum emission in the $M$-band originates close to the inner edge of the 
dusty molecular clouds closest to the AGN. The dust closest to the AGN 
probably has a high temperature, roughly equal to the sublimation temperature 
around $\sim$ 1500 K. 
This hot dust produces thermal emission whose spectral energy distribution peaks 
at near-infrared wavelengths, and also 
contributes significantly at 5 $\rm{\mu m}$ continuum emission 
(Neugebauer et al. 1987; Barvainis 1992; Kobayashi et al. 1993). 
The CO absorption lines would then be observed largely against continuum emission 
produced by this hottest dust nearest to the AGN, 
while the silicate absorption would 
occur largely against dust located somewhat more distant from it.

\section{Discussion}\label{sec:discussion}

Our spectrum toward IRAS~08572$+$3915 NW clearly separates the CO absorption 
into three velocity components; a cold component at the systemic velocity, 
a strong and warm component that is highly blueshifted ($-160$ km~s$^{-1}$), 
and a weaker, but much warmer component that is highly redshifted ($+$100 km~s$^{-1}$). 
From the population diagram of the observed CO absorption lines, 
we deduce that the blueshifted absorbing gas is dense ($> 10^{7}$ $\rm{cm^{-3}}$) 
and warm ($\sim$ 270 K) and has a CO column density of 
$N_{\rm{CO}} \sim 4.5 \times 10^{18}$ $\rm{cm^{-2}}$, which is 
equivalent to a $\rm{H_2}$ column density of $N_{\rm{H_2}} \sim 2.5 \times 10^{22}$ $\rm{cm^{-2}}$. 
The redshifted absorbing gas appears to be produced by even warmer and 
presumably denser gas. 
These high temperatures and high densities, as well as the small column lengths ($<$ 0.04 pc), 
indicate that these absorption components are in close proximity to the central engine. 

The physical properties that we derive are similar to the recent results 
from near- to mid-infrared spectroscopy of other ULIRGs. 
Lahuis et al. (2007) detected strong absorption bands of gaseous 
$\rm{C_2H_2}$, HCN, and $\rm{CO_2}$ toward deeply obscured (U)LIRG nuclei. 
Their data suggest the presence of warm (200--700 K) and 
dense ($> 10^7$ $\rm{cm^{-3}}$) gas. Armus et al. (2007) argued that 
this gas occupies only a small fraction of the nuclear region ($\sim$ 0.01 pc) 
near the mid-infrared continuum source. 
However, because all of their observations were made at much lower spectral resolution, 
they did not resolve individual velocity components. 
Our observations at high spectral resolution shed additional light on 
the geometrical structure and kinematics of the molecular clouds 
in one ULIRG and our conclusions may apply to others as well. 

The extremely deep silicate absorption in some ULIRGs 
requires their nuclear sources to be deeply embedded 
in optically and geometrically thick material (Levenson et al. 2007). 
Such absorptions together with pointlike infrared nuclear sources 
often are taken as signposts of the existence of AGNs (Imanishi, Dudley \& Maloney 2006). 
However, Spoon et al. (2006) discovered crystalline silicate features 
toward IRAS~08572$+$3915 in addition to the deep amorphous silicate feature, 
which can be a sign of high-temperature grain processing 
by the merger-triggered star-formation driven dust injection.

Various kinds of theoretical models for the material local to the AGN 
and obscuring it have been proposed. In all of them the central engines 
are obscured by dusty, optically and geometrically thick molecular clouds 
in toroidal structures; for example, a torus (Schartmann et al. 2005), 
a flared disk (Fritz et al. 2006), and discrete clouds (Elitzur \& Shlosman 2006). 
Krolik \& Begelman (1988) argued that the obscuring torus is composed of 
high-density molecular clouds. Nenkova et al. (2002) concluded that 
the obscuring torus is composed of a large number of small but dense molecular clouds. 
Indeed, the thermal dust emission from the torus has been spatially resolved 
by recent IR interferometric techniques and shown to have a clumpy or 
filamentary dust structure (Jaffe et al. 2004; Tristram et al. 2007). 
Additional evidence for the existence of the dust torus with many clumpy clouds 
comes from the detection of the silicate feature at $\sim$10 $\rm{\mu m}$ 
in the spectral energy distribution of AGNs. The recent discoveries of 
silicate emission from QSOs (Siebenmorgen et al. 2005; Hao et al. 2005), 
Seyferts (Hao et al. 2007), and LINERs (Sturm et al. 2005) by Spitzer provide 
strong evidence for a clumpy torus model. From a theoretical point of view, 
according to the 3-dimensional radiative transfer model of clumpy dust tori 
of AGN (Vollmer, Beckert \& Duschl 2004; Beckert \& Duschl 2004; 
H$\rm{\ddot{o}}$nig et al. 2006; H$\rm{\ddot{o}}$nig \& Beckert 2007; 
Nenkova et al. 2008), the properties of the individual self-gravitating clouds 
are speculated to be high density ($n \sim 10^{7-8}$ $\rm{cm^{-3}}$), 
large column density ($N_{\rm{H}} \sim 10^{24}$ $\rm{cm^{-2}}$), and 
small volume filling factor ($\sim$ 0.03). Wada (2007) investigated 
the possible origins of molecular absorption, based on 
a 3-dimensional hydrodynamic model of the inhomogeneous 
interstellar medium around the AGN. They demonstrated that the warm and 
dense molecular clouds, such as those discovered here via their CO absorption, 
can survive in the form of high-density clumps 
in the inhomogeneous interstellar matter around the AGN. 

Our high resolution spectroscopy has successfully detected 
warm dense gas with complex velocity structure, 
containing both highly blueshifted and highly redshifted components. 
Moreover, the CO spectrum suggests that the absorbing clouds have 
a large area covering factor ($>0.6$) but a small volume filling factor (thickness $<$ 0.04 pc). 
Because the velocity dispersions of the redshifted and blueshifted components 
are so large, it seems inconceivable that each exists in a single narrow sheet; 
hence each component must itself be composed of numerous well-separated thin sheets 
of dense gas that are detached from the continuum source. 

The redshifted and blueshifted components may trace the fueling 
of the central source and the ejection of material from it. 
However, most previous theoretical studies (e.g., Wada \& Tomisaka 2005) 
have concluded that the dominant velocity structure in the torus is 
systematic rotation and turbulence. For IRAS~08572$+$3915 NW 
such models are apparently inconsistent with our observed multiple, 
discrete velocity components. 
The picture that the absorbing clouds in the torus rotate around the AGN 
cannot reproduce the observed CO 
absorptions in IRAS~08572$+$3915 NW for the following reasons: 
(1) Absorption by gas clouds in orbital motion around the AGN 
should reproduce a broad absorption profile centered at 
the systemic velocity, because it will be make up 
of the integrated absorption of the entire line-of-sight velocity. 
The observed deep absorption lines with clear line separation 
into the blue-shifted and red-shifted components cannot be expected; 
(2) The redshifted absorption is extremely weak 
relative to the blueshifted absorption; and 
(3) The redshifted gas has much higher temperature than the blueshifted gas. 

We conclude that the velocity-shifted line components do not 
arise in a stable rotating torus-like gaseous structure, 
but are more likely associated primarily with ejection and 
injection of a dense and warm gas close to the AGN. 
Our results suggest that, at least in this case, 
the environment immediately surrounding the AGN does not appear 
to be as simple as that proposed in the unified scheme of AGNs.

\section{Summary}\label{sec:summary}

We have observed and analyzed a high-resolution $M$-band spectrum of the 
obscured AGN, IRAS~08572$+$3915 NW, which contains strong absorption lines of 
the fundamental ($v$=1$\leftarrow$0) band of CO. 
The main results and conclusions are as follows. \\
(1) Absorption lines of $^{12}\rm{C}^{16}\rm{O}$ $v$=1$\leftarrow$0 
are seen up to an excitation level of $J$=17. \\ 
(2) The absorbing gas has three velocity components: 
a cold component at the systemic velocity; 
a strong and warm component that is highly blueshifted ($-160$ km~s$^{-1}$); 
and a weaker, but much warmer component that is highly redshifted ($+$100 km~s$^{-1}$). \\
(3) Assuming LTE in the absorbing molecular gas, 
the temperatures of the systemic and blueshifted components are 
$23\pm1$ K and $325\pm5$ K, and the CO gas column densities are 
$N_{\rm{CO,cold}}=(5.74\pm0.06) \times 10^{17}$ $\rm{cm^{-2}}$ and 
$N_{\rm{CO,warm}}=(2.65\pm0.04) \times 10^{18}$ $\rm{cm^{-2}}$, respectively, 
for a covering factor of unity. For dense molecular gas these values correspond to 
$N_{\rm{H_2,cold}}=(3.19\pm0.03) \times 10^{21}$ $\rm{cm^{-2}}$ and 
$N_{\rm{H_2,warm}}=(1.47\pm0.02) \times 10^{22}$ $\rm{cm^{-2}}$. 
In the case of a covering factor of 0.6, which seems more likely based on 
optical depth arguments, the temperatures of these components are 
$24\pm1$ K and $273\pm2$ K, and the CO column densities are 
$N_{\rm{CO,cold}}=(1.98\pm0.01) \times 10^{18}$ $\rm{cm^{-2}}$ and 
$N_{\rm{CO,warm}}=(4.48\pm0.04) \times 10^{18}$ $\rm{cm^{-2}}$. 
These values correspond to 
$N_{\rm{H_2,cold}}=(1.10\pm0.01) \times 10^{22}$ $\rm{cm^{-2}}$ and 
$N_{\rm{H_2,warm}}=(2.49\pm0.02) \times 10^{22}$ $\rm{cm^{-2}}$, respectively. 
The redshifted component has a higher temperature ($\sim$ 700 K) 
than the blueshifted component and its column density is considerably less 
($N_{\rm{CO,hot}} \sim10^{17}$ $\rm{cm^{-2}}$, yielding $N_{\rm{H_2,hot}} \sim10^{21}$ $\rm{cm^{-2}}$). 
The H$_2$ column density associated with the cold gas at near systemic velocities 
may be considerably larger than the above values 
if the cold CO is largely in diffuse gas clouds. \\
(4) From an estimate of the critical density, 
we demonstrate that the molecular gas with warm temperature must be 
very dense ($n_{c}(\rm{H_2}) > 2\times10^7$ $\rm{cm^{-3}}$) and geometrically 
thin ($<$ 0.04 pc) along the line of sight, even if it is highly clumpy. \\ 
(5) The presence of discrete, highly blueshifted and highly redshifted 
absorption components close to the central engine does not fit the 
standard unified model of AGNs.

\bigskip

We acknowledge all of the staff and crew of the Subaru Telescope and 
NAOJ for their valuable assistance in obtaining this data and their 
continuous support for the construction and maintenance of IRCS. 
Particularly, we are grateful to H.~Terada and S.~Oya for technical 
support of these observations. We thank S.~Oyabu, S.~Matsuura, 
I.~Yamamura, K.~Wada, and T.~Kawaguchi for many fruitful discussions. 
T.~R.~G.'s research is supported by the Gemini Observatory, which is 
operated by the Association of Universities for Research in Astronomy, 
on behalf of the International Gemini Partnership of Argentina, 
Australia, Brazil, Canada, Chile, the United Kingdom, 
and the United States of America.





\end{document}